# Giant Magneto-electric Effect in the $J_{eff} = ½$ Mott Insulator, $Sr_2IrO_4$

S. Chikara[1,2], O. Korneta[1,2], W. P. Crummett[1,3], L.E. DeLong[1,2], P. Schlottmann[4] and G. Cao[1,2]*

[1]Center for Advanced Materials, University of Kentucky, Lexington, KY 40506

[2]Department of Physics and Astronomy, University of Kentucky, Lexington, KY 40506

[3]Science Division, Centre College, Danville, KY 40422

[4]Department of Physics, Florida State University, Tallahassee, FL 32306

Our magnetic, electrical, and thermal measurements on single-crystals of the $J_{eff} = 1/2$ Mott insulator, $Sr_2IrO_4$, reveal a novel giant magneto-electric effect (GME) arising from a frustrated magnetic/ferroelectric state whose signatures are: **(1)** a strongly enhanced electric permittivity that peaks near a newly observed magnetic anomaly at 100 K, **(2)** a large (∼100%) magneto-dielectric shift that occurs near a metamagnetic transition, and **(3)** magnetic and electric polarization hysteresis. The GME and electric polarization hinge on a spin-orbit gapping of 5d-bands, rather than the magnitude and spatial dependence of magnetization, as traditionally accepted.

It is commonly expected that iridates are more metallic and less magnetic than their 3d, 4d and 4f counterparts. The extended nature of the 5d orbitals leads to a broader 5d-bandwidth W and a reduced Coulomb interaction U such that $Ug(E_F) < 1$, where $g(E_F)$ is the density of states; the Stoner criterion therefore anticipates a metallic, paramagnetic state. In marked contrast, many iridates are magnetic insulators with exotic properties, such as the coexistence of a charge density wave and weak ferromagnetism (FM) in $BaIrO_3$ [1-3], anomalous "diamagnetism" in $Sr_3Ir_2O_7$ [10] and a novel $J_{eff} = 1/2$ Mott state in $Sr_2IrO_4$ [9,11]. Strong spin-orbit coupling (~0.3 - 0.4 eV, compared to ~20 meV in 3d materials) competes with other interactions in 5d materials to drive these exotic states.

In this paper, we report dielectric, magnetic, transport and thermal properties of single-crystal $Sr_2IrO_4$. We observe a novel giant magneto-electric effect (GME) characterized by a strongly peaked permittivity near a newly observed magnetic anomaly at 100 K, a large magneto-dielectric shift near a metamagnetic transition, and ferroelectric (FE) hysteresis. The GME and FE behavior hinge on strong spin-orbit coupling rather than the magnitude and spatial dependence of magnetization, contrary to current phenomenological models.

In $Sr_2IrO_4$, strong crystal fields split off 5d band states with $e_g$ symmetry (which play only a secondary role in physical properties), and $t_{2g}$ bands arise from $J = 1/2$ and $J = 3/2$ multiplets via strong spin-orbit coupling. A weak admixture of the $e_g$ orbitals downshifts the $J = 3/2$ quadruplet from the $J = 1/2$ doublet [9]. An independent electron picture anticipates a metallic compound, since the $Ir^{4+}$ ($5d^5$) ions provide four electrons to fill the lower $J_{eff} = 3/2$ bands, plus one electron to partially fill the $J_{eff} = 1/2$ bands.

Crucial to the present study of $Sr_2IrO_4$ is that even a modest U (~ 0.5 eV) can induce a Mott insulating gap Δ ~ 0.5 eV **[9, 11]** in the $J_{eff}$ = 1/2 band due to its narrow width (W = 0.48 eV), a circumstance that favors a recently predicted **[12]**, novel mechanism for a GME. The GME is an established feature of 3d multiferroics, but previously unobserved in 5d perovskites. Moreover, the traditional view is the GME depends only on the magnitude and spatial dependence of magnetization, whereas the new mechanism yields a FE polarization P that scales with an effective spin-orbit gap $Δ_s$ **[12]**. Our results indicate this new mechanism is realized in $Sr_2IrO_4$.

Synthesis and characterization of single-crystal $Sr_2IrO_4$ are described elsewhere **[7, 10]**. Measurements of specific heat C(T,H), magnetization M(T,H), ac susceptibility χ(T,H,ω), and electrical resistivity ρ(T,H) for T < 400 K were performed using either a Quantum Design PPMS or MPMS. High-temperature thermoelectric power S and ρ were measured from 9 K to 600 K. The complex permittivity ε(T,H,ω) = ε' + iε" was measured using a 7600 QuadTech LCR Meter with 10 Hz ≤ ω ≤ 2 MHz and H < 12 T. The electric polarization was measured using a Radiant Precision Premier II polarimeter.

The complex behavior of $Sr_2IrO_4$ demands a careful comparison of new single-crystal data for M(T), C(T), ρ(T) and S(T), as shown in **Fig. 1**. Our results agree with previous magnetic data for $Sr_2IrO_4$ which revealed weak FM order below $T_C$ = 240 K **[4-8]** and a low-field metamagnetic transition resulting in a small saturation moment $μ_s$ < 0.13 $μ_B$/Ir (sample-dependent) along the easy **a-**axis **[7]**. The Curie-Weiss temperature $θ_{cw}$ = +236 K extrapolated from the inverse susceptibility $Δχ^{-1}$ (Δχ = χ(T) - $χ_o$, where $χ_o$ is a T-independent contribution) confirms FM exchange coupling over the range 270 < T < 350 K (**Fig. 1a**). Arrott plots also corroborate weak FM order at $T_C$ ≈ 240 K (**Fig. 1d** inset).

On the other hand, C(T<11 K) is predominantly proportional to $T^3$ at $\mu_o H = 0$ and 9 T (**Fig. 1b** inset), due to a Debye-phonon and/or magnon contributions from an *AFM* ground state, in apparent conflict with the weak *FM* behavior suggested by **Fig. 1a**. Indeed, the field-shift [C(T,H)-C(T,0)]/C(T,0) ~ 16% at 9 T indicates a significant magnetic contribution to C(T), and suggests that a competition between AFM and FM exchange produces a low-temperature C(T) consistent with AFM magnons. The relative strength of AFM and FM interactions clearly shifts to drive different magnetic states at high and low T, to which we return below. A tiny specific heat anomaly ($|\Delta C| \sim 4$ mJK/mole K) is observed at $T_C$, indicating a very small entropy change (**Fig. 1b**) (complete absence of a C(T) anomaly was previously noted [8]).

Remarkably, we find no anomaly at $T_C$ in $\rho(T)$ and S(T) (**Fig. 1c** and **1d**), which is perplexing, since the M(T≈$T_C$) anomaly is robust and indicates long-range magnetic order. (Note the transport properties of $Sr_2IrO_4$ are extremely sensitive to oxygen content, but the magnetism is not [8,14].) The absence of a phase transition signature in $\rho$(T≈$T_C$) and S(T≈$T_C$) could reflect a spin-glass state. The real part $\chi'$ of the ac magnetic susceptibility shown in **Fig. 2a** exhibits no sharp, $\omega$-dependent peak near $T_C$ which would *clearly signal* spin glass behavior. We note that $\chi'$(T) displays a pronounced peak near 135 K and a smaller peak near 85 K, both highly sensitive to dc magnetic field (**Fig. 2a**).

A key feature in **Figs. 2b** and **2c** is the newly observed magnetic anomaly below $T_M \approx$ 100 K, whose location is very sensitive to H, and which we argue may be a result of *gradual spin canting*. $Sr_2IrO_4$ crystallizes in the reduced tetragonal space-group $I4_1/acd$ [4, 5] due to a rotation of the $IrO_6$-octahedra about the **c-**axis by ~11°, which removes the I4/mmm inversion center existing between the Ir ions along the (100) and (010) directions (inset, **Figs. 2b** and **2c**) **[4]**. The rotation increases from 11.36° at room temperature to 11.72° at 10 K **[5]**, corresponding to a reduction of the Ir-O-Ir bond angle from 157.28° to 156.56°, respectively, and accompanies a growth of the **c**-axis lattice parameter **[4]**. Moreover, a strong temperature dependence of bending modes associated with the Ir-O-Ir bond angle was recently observed **[13]**, which, in turn, influences the magnetic exchange interaction. The data in **Fig. 2** suggest that $T_M$ marks a crossover of the dominant exchange coupling from FM to AFM (discussed earlier), with a reduction of the Ir-O-Ir bond angle. Increased spin canting, or an AFM state, at low T is also consistent with the downturn in M, substantial rise in $\rho(T<T_M)$, and $C(T) \propto T^3$ seen at low T in **Fig. 1**.

Spin canting, the T-dependent Ir-O-Ir bond angle, and loss of inversion symmetry also influence the dielectric behavior and potential for FE order that is strongly dependent on crystal symmetry. Indeed, the magnetic anomaly at $T_M$ is closely linked to the dielectric response, as shown in **Figs. 3a** and **3b**, where M(T) is compared to the real parts $\varepsilon_c'(T)$ and $\varepsilon_a'(T)$ of the **c**-axis and **a**-axis dielectric constants, respectively. Two major features emerge: **(1)** both $\varepsilon_c'(T)$ and $\varepsilon_a'(T)$ rise by up to ***one order of magnitude*** and peak near $T_M$, similar to $La_2CuO_4$ **[15]**. ($\varepsilon_a(T)$ is loss-dominated above $T_M$, therefore we focus only on $\varepsilon_c'(T)$ in the discussion that follows.) This strong enhancement of $\varepsilon_c'(T)$ is much larger than that exhibited by well-known magneto-electrics such as $BaMnF_4$ **[16]**, $BiMnO_3$ **[17]**, $HoMnO_3$ and $YMnO_3$ **[18]**. **(2)** The peak in $\varepsilon_c'(T \approx T_M)$ separates two regions, I and II, as marked in **Fig. 3a**. The weak frequency dependence of $\varepsilon_c'(T,\omega)$ in low-T Region I is typical of a ferroelectric, whereas the stronger frequency dependence of $\varepsilon_c'(T,\omega)$ in higher-T Region II suggests a relaxor mechanism **[19]**, which is traditionally attributed to disorder and impurities. Alternatively, the sharp peak accompanied by strong frequency dispersion could signal a novel frustrated or disordered magnetic/FE state corresponding to the shaded area in **Figs. 3a** and **3b**.

The vast majority of known magneto-electrics and multiferroics are 3d-based compounds **[15-24]**, whereas there are no known examples of ferroelectric 5d materials, so it is of interest to determine if $Sr_2IrO_4$ is indeed a ferroelectric. Electric polarization hysteresis is certainly observed in $Sr_2IrO_4$ as shown in **Fig. 3c**, suggesting the existence of some type of FE state at low T (A detailed description of evidence for the FE state will be published elsewhere). Furthermore, a magneto-dielectric shift $\Delta\varepsilon_c'(H)/\varepsilon_c'(0)$ is also anticipated and observed in $M_c(H)$ near the metamagnetic transition field $H_c$ (**Fig. 4**). (The negligible magnetoresistance in $Sr_2IrO_4$ at H up to 12 T **[7]** suggests that $\Delta\varepsilon_c'(H)/\varepsilon_c'(0)$ is an intrinsic effect.) However, *we do not observe* $\Delta\varepsilon_c'(H) \propto M^2$ as conventionally expected **[17]**, and M ($< 0.1$ $\mu_B$/Ir) is exceptionally weak compared to known multiferroics (e.g., M $\approx$ 6 $\mu_B$/f.u. for $TbMnO_3$ **[23]**).

Although the GME in $Sr_2IrO_4$ is unconventional, it can be understood as a unique manifestation of a recently formulated microscopic mechanism for magneto-electrics with strong spin-orbit coupling; this novel approach yields P proportional to an effective spin-orbit gap $\Delta_s$ rather than the magnitude and spatial dependence of the magnetization **[12]**.

In light of all results presented above, it is suggested that $T_M$ defines a drastic change in the coupling between the magnetic and dielectric response, according to the following scenario: In Region II, the strong competition between FM and AFM exchange couplings promotes frustrated or incommensurate magnetic order. A T-dependent magneto-elastic coupling may give rise to a soft lattice mode, as indicated by optical data [13] and the weak frequency dependence of χ' near 135 K (**Fig. 2a**). The Ir-O-Ir bond angle (**Fig. 3a**) decreases with decreasing T, strengthening the AFM exchange coupling until, near $T_M$, the AFM coupling becomes dominant, and spins are "locked in" with a stiffened lattice in Region I. This scenario explains **(a)** $C(T) \propto T^3$ below 11 K (**Fig. 1b**), **(b)** the low-T downturn in M (**Figs. 1a** and **2b**), **(c)** the rise of ρ(T) below $T_M$ (**Fig. 1c**), **(d)** the reduction of the frequency dependence of $\varepsilon_c'(T,\omega)$ in Region I (**Fig. 3a**), and **(e)** the reduction of the magneto-dielectric effect $\Delta\varepsilon_c'(H)/\varepsilon_c'(0)$ from 100% at 50 K, to only 21% at 110 K (**Fig. 4a**).

In summary, a dominant spin-orbit coupling in $Sr_2IrO_4$ shifts the balance of competing magnetic, dielectric and lattice energies, generating a novel type of GME that is not dependent on the magnetization, but nevertheless is intimately linked with the complex magnetic order emerging from an exotic Mott insulating state. We expect further examples of exciting new type of GME and multiferroics to be found in other 5d Mott insulators.

This work was supported by NSF through grants DMR-0552267 and EPS-0814194. LED was supported by DoE through grant DE-FG02-97ER45653. PS was supported by the DoE through grant DE-FG02-98ER45707.

*Corresponding author; email:cao@pa.uky.edu

**Figure Captions:**

**Fig.1. (a)** Field-cooled magnetization M(T) and inverse susceptibility $\Delta\chi^{-1}$(T) (right scale) at applied field $\mu_oH$ = 0.2 T, **(b)** Specific heat C(T), **(c)** resistivity $\rho$(T) and **(d)** thermoelectric power S(T) for $Sr_2IrO_4$ as functions of temperature T $\leq$ 600 K. Except for C(T) (**H$\perp$c**), all properties are measured for **H** along both **a-** and **c-**axis. **Fig. 1b inset**: C(T) vs $T^3$ for T < 11 K. **Fig. 1d inset**: Arrott plot ($M^2$ vs $\mu_oH/M$) shows a FM transition at $T_C \approx$ 240 K. Note: no anomaly is seen in $\rho(T \approx T_C)$ and $S(T \approx T_C)$.

**Fig. 2. (a)** Real part of ac susceptibility $\chi$'(T,$\omega$) along the **a-**axis at $\mu_oH$ = 0.1 T and frequencies $\omega$ = 0.8 and 80 Hz. **(b)** dc magnetizations $M_a$(T) and **(c)** $M_c$(T) for various magnetic fields H. Note the H-dependent anomaly at $T_M$ (arrows). **Fig. 2a inset:** $\chi$'(T) for $\omega$ = 8 Hz at $\mu_oH$ = 0.01 and 0.1 T. **Figs. 2b-2c inset:** Rotation scheme for $IrO_6$ octahedra.

**Fig. 3. (a)** Real part of the **c-**axis dielectric constant $\varepsilon_c$'(T) for representative frequencies $\omega$ (left scale), and **c**-axis magnetization $M_c$(T) (right scale); **Fig. 3a inset:** Schematic change of O-Ir-O bond angle from Region I to Region II. **(b)** Real part of the **a-**axis dielectric constant $\varepsilon_a$'(T) for representative $\omega$ (left scale), and **a-**axis $M_a$(T) (right scale). **(c)** Electric polarization P vs voltage V for temperature T = 13.3 K (low V, left scale) and 4.3 K (high V, right scale).

**Fig. 4. (a)** Magneto-electric effect $\Delta\varepsilon_c'(H)/\varepsilon_c'(0)$ along **c**-axis at temperatures T = 50 K and 110 K versus applied field H for a few representative frequencies $\omega$ and $\mu_o H \leq 10$ T applied along **c**-axis; Right scale: the **c**-axis $M_c(H)$ vs. H at 1.7 K; **(b)** the **c**-axis $M_c(H)$ at various T. Note parallel behavior of $\varepsilon_c$ and $M_c$ near the metamagnetic transition field, $H_c$.

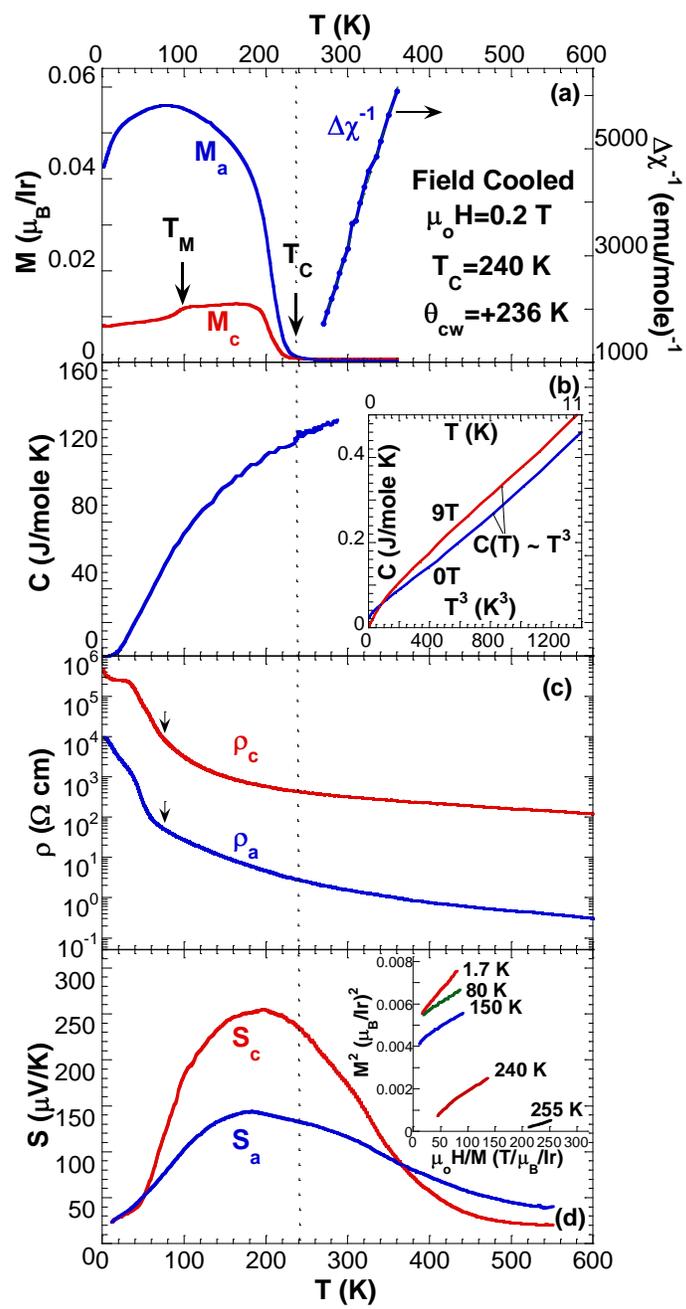

Fig. 1

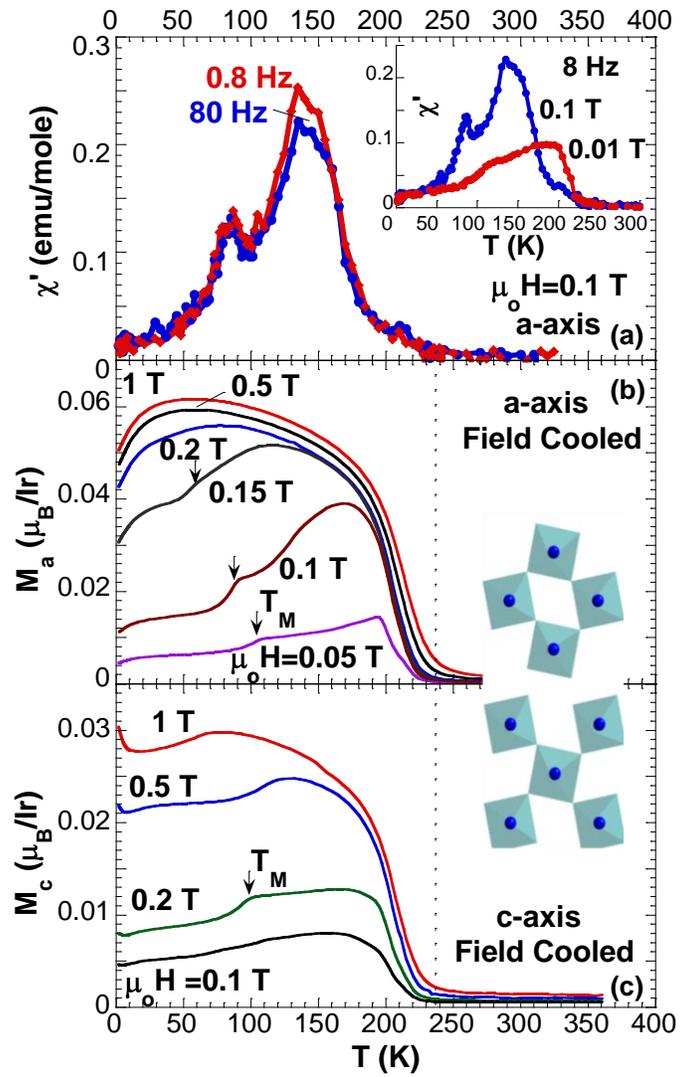

Fig. 2

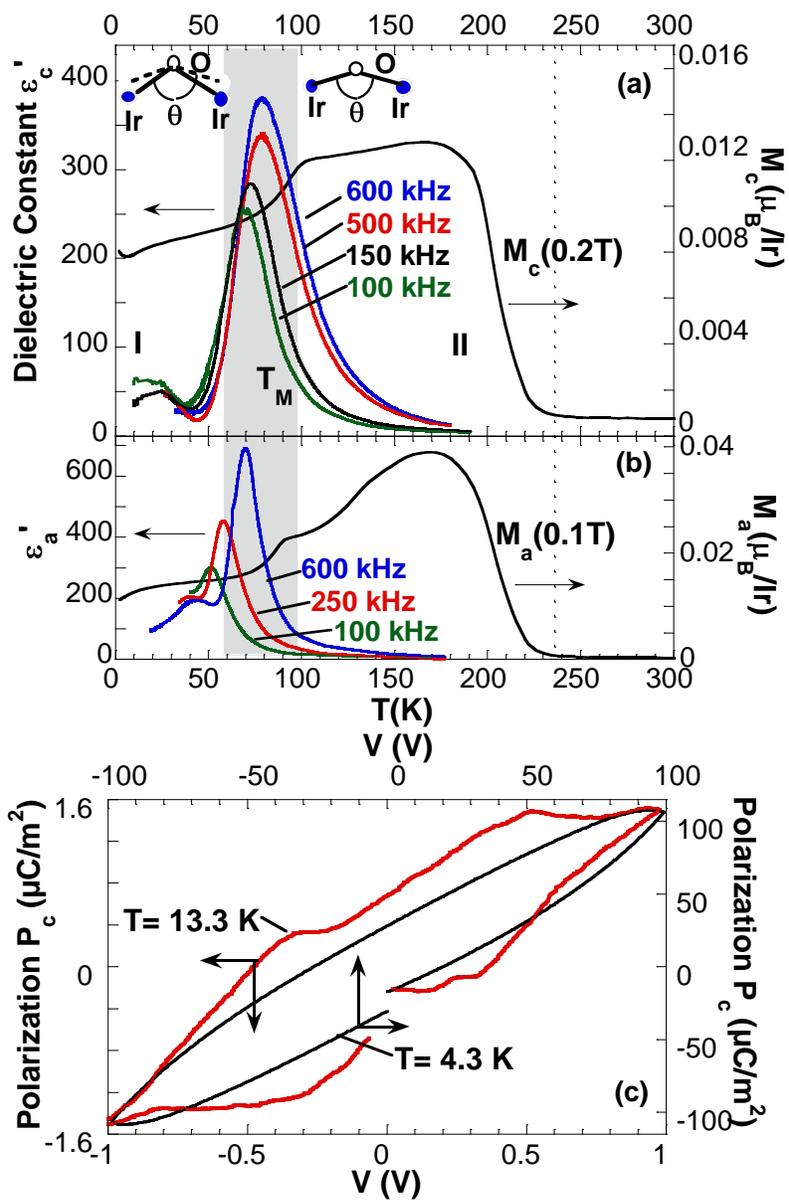

Fig. 3

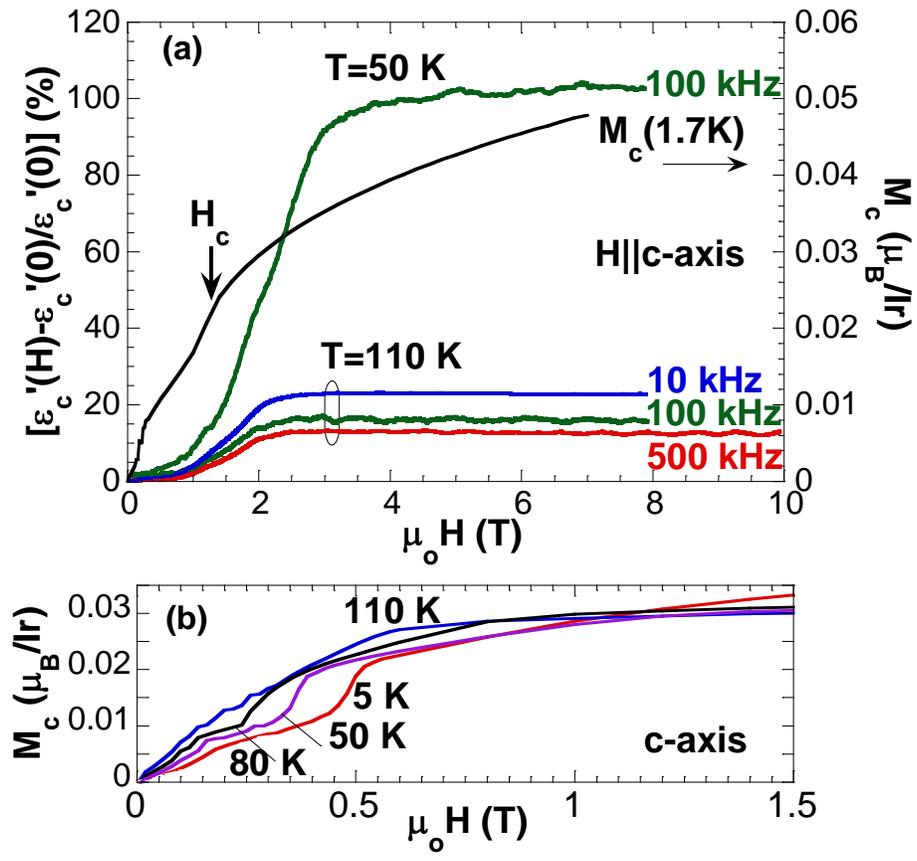

Fig.4